\documentclass{article}
\pdfoutput=1
\usepackage{spconf,amsmath,graphicx}

\usepackage{cite}
\usepackage{multirow}
\usepackage{multicol}
\usepackage{booktabs}
\usepackage{adjustbox}
\usepackage{float}
\usepackage{amsmath,bm,soul}
\usepackage[table]{xcolor}
\usepackage{arydshln}
\usepackage{fancyhdr}

\usepackage{etoolbox}

\title{Silver Standard Masks for Data augmentation Applied to Deep-Learning-Based Skull-Stripping}
\name{Oeslle Lucena$^{\star}$, Roberto Souza$^{\star \dagger}$, Let\'{i}cia  Rittner$^{\star}$, Richard Frayne$^{\dagger}$, Roberto Lotufo$^{\star}$}
\address{$^{\star}$ University of Campinas, Campinas, S\~ao Paulo, Brazil \\
    $^{\dagger}$ University of Calgary, Calgary, Alberta, Canada 
    } %
\fancypagestyle{firstpage}{%
\chead{This work has been submitted to the IEEE for possible publication. Copyright may be transferred without notice, after which this version may no longer be accessible.}
}

\begin{document}
%
\maketitle
\thispagestyle{firstpage}
\begin{abstract}
The bottleneck of convolutional neural networks (CNN) for medical imaging is the number of annotated data required for training. Manual segmentation is considered to be the ``gold-standard''. However, medical imaging datasets with expert manual segmentation are scarce as this step is time-consuming and expensive. We propose in this work the use of what we refer to as silver standard masks for data augmentation in deep-learning-based skull-stripping also known as brain extraction. We generated the silver standard masks using the consensus algorithm Simultaneous Truth and Performance Level Estimation (STAPLE). We evaluated CNN models generated by the silver and gold standard masks. Then, we validated the silver standard masks for CNNs training in one dataset, and showed its generalization to two other datasets. Our results indicated that models generated with silver standard masks are comparable to models generated with gold standard masks and have better generalizability. Moreover, our results also indicate that silver standard masks could be used to augment the input dataset at training stage, reducing the need for manual segmentation at this step.
\end{abstract}
\section{Introduction}
\label{sec:intro}
Segmenting brain tissues from non-brain tissues is known as skull-stripping (SS) or brain extraction. Skull stripping is usually an initial step for many other types of image processing in brain MR images analyses, for instance in segmenting tissue types~\cite{de2010accuracy}, 
monitoring the development or aging of the brain~\cite{blanton2004gender}, and in determining abnormal volumes and shapes across many brain disorders~\cite{petrella2003neuroimaging,hutchinson2000structural}.

Since the work of Krizhevsky et al.~\cite{krizhevsky2012imagenet} at the ImageNet contest in 2012, deep learning (DL) approaches have been extensively used, becoming a commonly used algorithm to solve many problems in the medical imaging  field. For instance, DL was successfully applied in brain anatomy segmentation~\cite{de2015deep}, and brain tumor segmentation~\cite{pereira2016brain}. Regarding medical image segmentation, fully convolutional networks (FCNs)~\cite{long2015fully} have been employed to this task. FCNs performs optimally in identifying local and global features within a computationally efficient training period~\cite{long2015fully,ronneberger2015u}.

In medical imaging datasets, manual segmentation performed by experts is usually considered the ``gold standard" masks. Nevertheless, manual segmentation is a time-consuming and expensive task~\cite{greenspan2016guest}, because the rater has to manually delineate each single voxel for a 3D volume, taking up to hours to correctly segment a human brain. Moreover, the manual segmentation guidelines vary among experts~\cite{warfield2004simultaneous}, thus, suffering from intra- and inter-rater variability~\cite{asman2011robust}. To this end, new ways to generate automatic labeled data are investigated for medical imagery analysis, such as the consensus methods that through agreement algorithms combine automatic methods to generate what we refer to as ``silver standard" masks~\cite{warfield2004simultaneous,asman2011robust,rex2004meta}. Consensus methods are very robust. Rex et al.~\cite{rex2004meta} compared the results of their consensus algorithm combining automatic methods and obtained a higher agreement rate than different segmentations done by two different experts. Souza et al.~\cite{souza2017open} released the \emph{CC-359} dataset which has silver standards masks generated by the consensus algorithm Simultaneous Truth and Performance Level Estimation (STAPLE)~\cite{warfield2004simultaneous}. Further, they suggested using consensus masks for convolutional neural networks (CNNs), but as far as we know, this usage is yet to be validated. 

Usually, training CNNs from scratch requires a large amount of labeled data. Nonetheless, datasets with gold standard masks are relatively small.
We present in this paper a data augmentation approach for deep-learning-based skull-stripping, which we validate the use of silver standards masks for the CNN training. The goal of this work is to show that silver standards masks have similar performance when compared to the gold standard masks for supervised CNN training. If demonstrated, it would then be possible to automatically generate labeled data and augment the number of training data, making it possible to enlarge datasets for large-scale analysis. We chose to validate our approach on SS because of its direct impact on the clinical/research analyses after the extraction of the brain and due to its time-consuming manual delineation. The silver standard masks are generated using the consensus algorithm STAPLE, which is an expectation-maximization algorithm that considers a collection of segmentations to compute a probabilistic estimate of the true segmentation. 

This work is organized as follow: In Section~\ref{sec:mat} we present all the materials and describe the methods for our experiments. The results and discussion of our analysis are detailed in Section~\ref{sec:res}. Finally, the conclusions of our study are presented in Section~\ref{sec:conc}.

\section{Materials and Methods}
\label{sec:mat}
\subsection{Datasets}
We used three datasets for our analysis. The first datset was the LONI Probabilistic Brain Atlas (LPBA40) set composed of 40 T1-weighted volumes from healthy subjects and their corresponding manually labeled brain masks~\cite{shattuck2008construction}. The second one was \emph{CC-359} - a public dataset composed of 359 subjects T1 image volumes~\cite{souza2017open}. The \emph{CC-359} includes the original volumes, the consensus masks generated for all subjects using the STAPLE algorithm, and twelve manual segmentations.
In this work, we only used the twelve image volumes with manual segmentation, which we refer to as \emph{CC-12}. Finally, we used the first two discs of the Open Access Series of Imaging Studies (OASIS) dataset which consists of T1-weighted volumes from 77 subjects~\cite{marcus2007open}. 

\subsection{STAPLE Silver-standard Masks}
\begin{figure}[!ht]
\begin{center}
\includegraphics[width=0.42\textwidth]{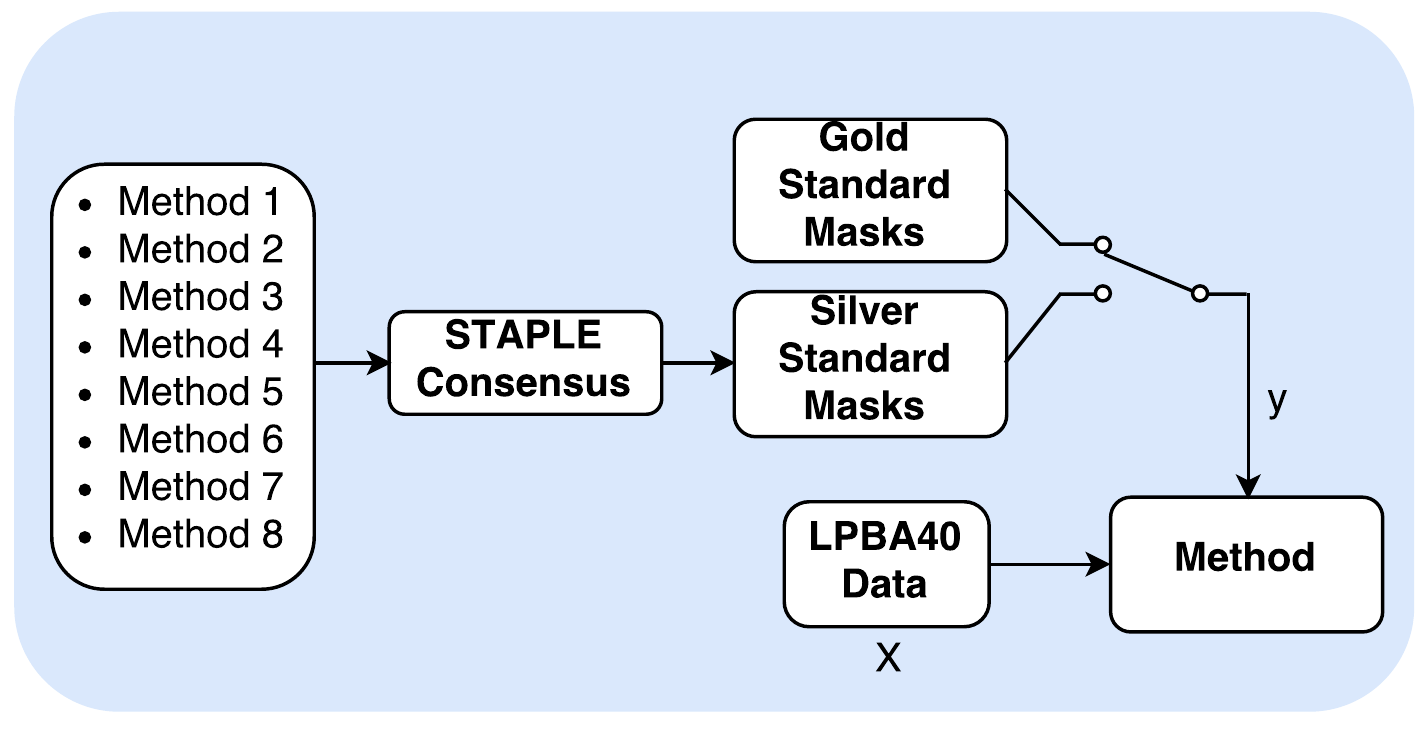}
\caption{\label{fig:input} Procedures adopted to conduct the first experiment. At the training stage in our method pipeline we used the gold and silver standard masks.
We used STAPLE consensus to generate the silver standard }
\end{center}
\end{figure}

The STAPLE output, which is a probability mask, is thresholded at $0.5$ to generate the silver standard masks. To this end, the method receives as input the masks resulting from the eight automatic non-deep learning SS methods~(Figure~\ref{fig:input}). These eight methods are the same ones used in Souza et al.~\cite{souza2017open}.
STAPLE is a consensus forming algorithm that uses an expectation-maximization algorithm to estimate the hidden true segmentation as a probabilistic mask. The algorithm considers a collection of segmentations and computes a probabilistic estimate of the true segmentation and a measure of the performance level represented by each segmentation~\cite{warfield2004simultaneous}. 

\begin{figure}[!ht]
\begin{center}
\includegraphics[width=0.45\textwidth]{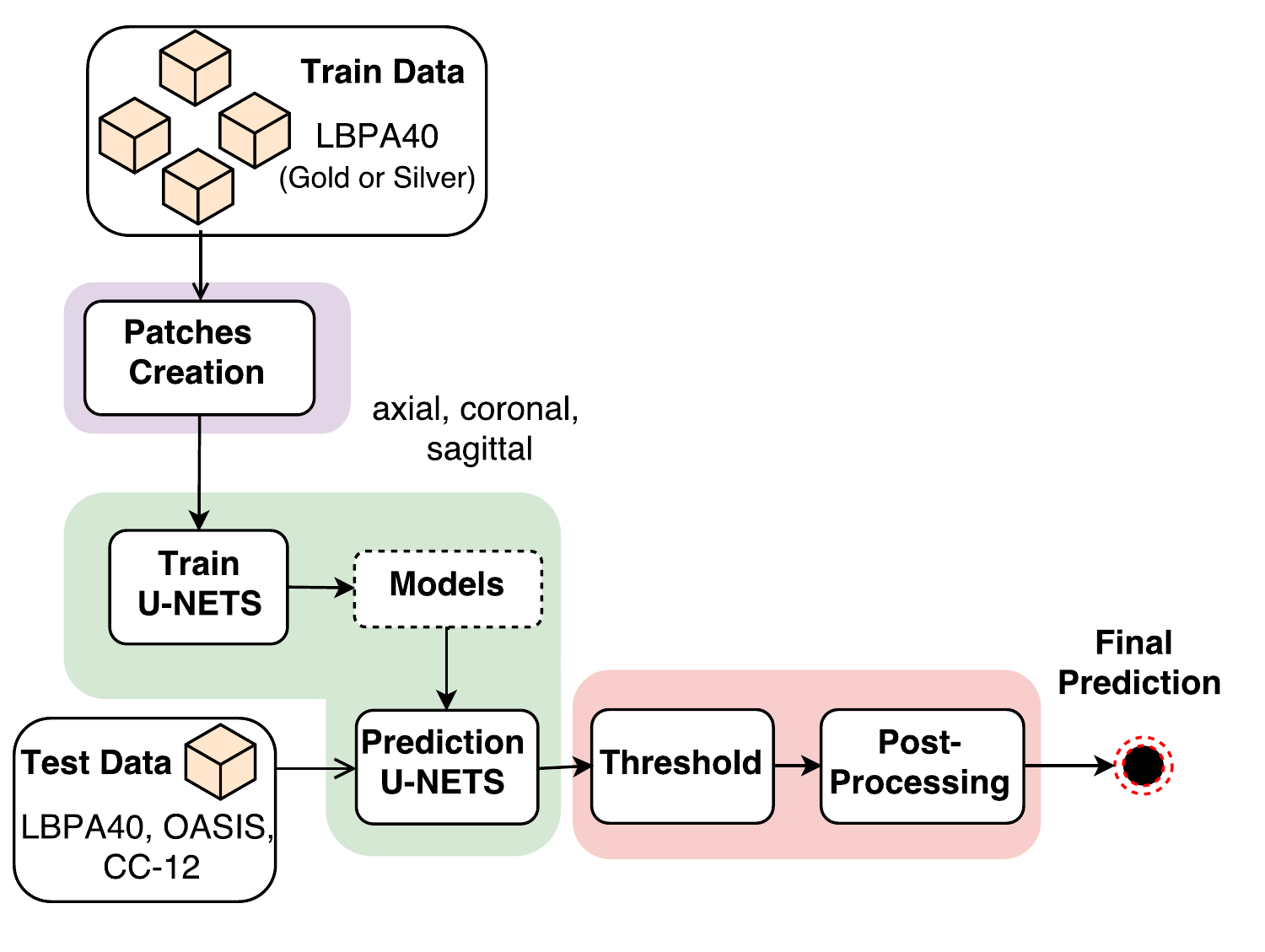}
\caption{\label{fig:ss-pipeline} The pipeline consists of three stages: patch creation (purple), deep segmentation (green), and threshold/post-processing (red).}
\end{center}
\end{figure}

\subsection{Metrics and Implementation Environment}
The metrics used to evaluate the segmentations were: Dice coefficient, sensitivity, specificity, Hausdorff distance, and mean symmetric surface-to-surface distance. These metrics are commonly used in SS analysis~\cite{iglesias2011robust,souza2017open}. 
Two-sided paired $t$-tests were used to assess statistical differences in the evaluation metrics. A $p<0.05$ was deemed to be significant. Our method was built based on a public Keras~\footnote{https://keras.io/} implementation of a 2D FCN U-Net~\footnote{https://github.com/jocicmarko/ultrasound-nerve-segmentation} and the full code of our implementation will be made available before publication. 

\begin{table*}[!ht]
\centering
\caption{\label{tab:results} Overall analysis against gold standard masks for the LPBA40, \emph{CC-12}, and OASIS datasets when trained with gold standard masks and silver standard masks from the LPBA40 dataset. For Dice, sensitivity and specificity higher values are better; and for the distances lower values are better. For Dice, sensitivity and specificity higher values are better; and for the distances lower values are better.}
\resizebox{.85\textwidth}{!}{
\begin{tabular}{cccccc}
\toprule
\multirow{2}{*}{\textbf{Training Masks}}  & \multicolumn{5}{c}{\textbf{Metrics}}\\
\cmidrule{2-6}
 & \textbf{Dice} (\%) & \textbf{Sensitivity} (\%) & \textbf{Specificity} (\%) & \textbf{Hausdorff} (mm) & \textbf{Mean} (mm) \\
\midrule
\multicolumn{6}{c}{\textbf{LPBA40}}\\
\midrule
\textbf{Gold} 
& $ 96.111 \pm 0.616 $ & $ 97.795 \pm 1.575 $ & $ 98.979 \pm 0.389 $ &
$ 13.086 \pm 3.585 $ &
$ 0.066 \pm 0.015 $ \\
\textit{p-value} & $\boldsymbol{0.005}$  & $0.573$  & $\boldsymbol{0.04}$  & $0.282$  & $\boldsymbol{0.003}$
\\
\textbf{Silver} 
& $ 95.793 \pm 0.931 $ & $ 97.715 \pm 1.274 $ & $ 98.876 \pm 0.517 $ &
$ 12.494 \pm 3.996 $ &
$ 0.075 \pm 0.024 $ \\ 
\midrule
\multicolumn{6}{c}{\textbf{CC-12}}\\
\midrule
\textbf{Gold} 
& $ 85.781 \pm 10.05 $ & $ 78.923 \pm 15.522 $ & $ 99.633 \pm 0.316 $ &
$ 16.402 \pm 7.118 $ &
$ 0.496 \pm 0.591 $ \\
\textit{p-value} & $\boldsymbol{5.318e-11}$  & $\boldsymbol{5.029e-14}$  & $\boldsymbol{5.453e-4}$  & $0.680$  & $\boldsymbol{1.632e-6}$
\\
\textbf{Silver} 
& $ 88.873 \pm 7.876 $ & $ 84.254 \pm 13.364 $ & $ 99.52 \pm 0.441 $ &
$ 17.106 \pm 12.669 $ &
$ 0.323 \pm 0.381 $ \\ 
\midrule
\multicolumn{6}{c}{\textbf{OASIS}}\\
\midrule
\textbf{Gold} 
& $ 88.009 \pm 5.248 $ & $ 79.682 \pm 8.37 $ & $ 99.695 \pm 0.282 $ &
$ 15.192 \pm 2.79 $ &
$ 0.334 \pm 0.19 $ \\
\textit{p-value} & $\boldsymbol{1.601e-39}$  & $\boldsymbol{8.364e-49}$  & $\boldsymbol{1.208e-14}$  & $0.830$  & $\boldsymbol{1.709e-21}$ 
\\
\textbf{Silver} 
& $ 89.329 \pm 4.382 $ & $ 81.968 \pm 7.349 $ & $ 99.591 \pm 0.386 $ &
$ 15.15 \pm 4.234 $ &
$ 0.29 \pm 0.162 $ \\ \bottomrule
\end{tabular}
}
\end{table*}

\subsection{Method}
The pipeline of our method uses parallel CNN models, one for each image plane (axial, coronal, and sagittal). The key idea was to perform 2D segmentation on a slice-by-slice basis for each image volume and repeat for the other two orthogonal orientations. 3D segmentation was then done by reconstruction through the concatenation of the 2D predictions. We used the 2D FCN U-Net which is a U-shaped network (contracting path, left side; expansive path, right side) composed of 23 convolutional layers~\cite{ronneberger2015u}. For the CNN parameters, we adopted the Adam optimizer with the configurations provided by the authors but changing the learning rate to $10^{-5}$. Moreover, the negative of the Dice coefficient was used as the loss function. The pipeline is presented in Figure~\ref{fig:ss-pipeline} with three major stages: 1) patch creation, 2) deep segmentation, and 3) threshold and post-processing. 

In the patch creation stage, we first normalized the image volumes to be in the same range (0 to 1000); a range chosen to ensure sufficient dynamic range and to minimize data storage limitations. Secondly, to increase the number of input data, we extracted five patches of size $64 \times 64$ from each slice that contained brain voxels, non-zero values in the corresponding mask. The deep segmentation stage consists of both training and prediction steps of the CNNs. In the training step, the parallel CNNs were trained with extracted patches as input, and the prediction step consisted of inferring one prediction image volume for each model using the whole image as input. In the threshold step, we set the value in the voxel to 1.0 if the average probability from the predictions of the three models is $\geq0.5$. Otherwise, the voxels were set to $0.0$. Lastly, the final prediction is obtained after a post-processing step where the largest connected component was preserved and smaller components were filtered out.

\section{Experiments and Results}
\label{sec:res}
Our comparison of the silver standard and gold standard masks was conducted with two experiments: Experiment 1) comparison/validation of silver and gold standard masks for CNN training (leading to the generation of silver and gold standard models, respectively), Experiment 2) generalization of silver and gold standard models to other datasets. For both experiments, we performed a five-fold cross-validation using the LPBA40 dataset to generate the models, which we initially generated its silver standard masks using STAPLE. (The image volumes of OASIS and the \emph{CC-12} datasets were only used in the second experiment.) The first experiment consisted of separately training the image volumes of the LPBA40 dataset with gold standard masks and then with the silver standard masks. We compared our results against the gold standard masks of the LPBA40 (Figure~\ref{fig:input}). The second experiment consisted of the models provided in experiment one, to see how well they predicted SS in the OASIS and the \emph{CC-12} data. We compared these predictions against the relevant gold standard masks. Table~\ref{tab:results} summarizes the overall analysis highlighting all metrics where a statistically significant difference ($p<0.05$) was observed.


The performance of the silver standard models, (models using silver standard masks) \color{black}, were comparable to the performance of gold standard models, (models using gold standard masks at the training stage) in experiment one~(Table~\ref{tab:results}). The gold standard models had better performance on Dice coefficient, which is the first metric observed to evaluate an optimal segmentation. However, the guidelines used for the manual segmentation are the same in the LPBA40. Thus, this bias improved gold standard model performance. Silver standard perfomance was comparable to gold standard models when tested against the \emph{CC-12} and OASIS data in the second experiment~(Table~\ref{tab:results}). The silver standard models were better than gold statrd models, likely because manual segmentations are biased towards their own guidelines while silver standard masks take advantage of the consensus approach reducing the intra-rater variability. Moreover, our results suggest that silver standard models have better generalization than gold standard models in a robust~(\emph{CC-12} subset) and manually corrected gold standard masks~(OASIS dataset).   

\section{Conclusions}
\label{sec:conc}
In this work, we have proposed the usage of silver standard masks for data augmentation in a deep-learning-based skull-stripping. The overall analysis indicated that silver standard models are comparable to gold standard models but generalize better due to consensus method, likewise STAPLE, reduce the intra-rater variability. Therefore, our results also indicate that silver standard masks could be used to augment the input dataset at training stage, reducing the need for manual segmentation at this step.

\section*{Acknowledgments}
Oeslle Lucena thanks FAPESP (2016/18332-8). Roberto A. Lotufo thanks CNPq (311228/2014-3),  Leticia Rittner thanks CNPq (308311/2016-7), Roberto Souza thanks the NSERC CREATE I3T foundation. Richard Frayne is supported by the Canadian Institutes for Health Research (CIHR, MOP-333931).




\bibliographystyle{IEEEbibabrv}
\bibliography{IEEEabrv,refs}

\end{document}